\documentclass{article}
\usepackage[super,compress]{cite}
\usepackage{url}

\usepackage{amsmath}

\usepackage{amssymb}

\usepackage{graphicx}

\begin{document}

\title{NEUTRAL DYNAMICS AND CELL RENEWAL OF COLONIC CRYPTS IN HOMEOSTATIC REGIME.}


\author{A.J.Fendrik\\ \small \it Instituto de Ciencias, Universidad Nacional de General Sarmiento.\\ \small \it   J.M.Gutierrez 1150, (1613) Los Polvorines, Buenos Aires, Argentina.\\ \small  \it Consejo Nacional de Investigaciones Cient\'{\i}ficas y T\'ecnicas- Buenos Aires, Argentina. \and L.Romanelli \\ \small \it Instituto de Ciencias, Universidad Nacional de General Sarmiento.\\     \small \it J.M.Gutierrez 1150, (1613) Los Polvorines, Buenos Aires, Argentina.\\ \small \it Consejo Nacional de Investigaciones Cient\'{\i}ficas y T\'ecnicas- Buenos Aires, Argentina. \and E.Rotondo \\ \small \it Instituto de Ciencias, Universidad Nacional de General Sarmiento.\\     \small \it J.M.Gutierrez 1150, (1613) Los Polvorines, Buenos Aires, Argentina.}

\maketitle

\begin{abstract}

The self renewal process in colonic crypts is the object of several studies. We present here a new compartment model with the following characteristics: a) We distinguish different classes of cells: stem cells, 6 generations of transit amplifying cells and the differentiated cells.  b) In order to take into account the monoclonal character of the crypts in homeostatic regime we include symmetric divisions of the stem cells. We first consider the dynamic differential equations that describe the evolution of the mean values of the populations but the small observed value of the total number of cells involved plus the huge dispersion of experimental 
data found in the literature leads us to study the stochastic discrete process. This analysis allows us to study fluctuations, the neutral drift that leads to monoclonality and the effects of the fixation of mutant clones.

\end{abstract}

\section{\label{sec:Intro} Introduction}
The cell renewal dynamics in  Lieberk\"{u}hn crypts located in the intestinal epithelium have been the object of many experimental and theoretical 
studies. This is due to the fact that the anomalies and disruptions in their functioning are related with the development of colonic cancer as well as to clarify the rhythm of cellular divisions in the  homeostatic regime to ensure the tissues renewal.

Indeed, in multicellular organisms cell divisions from stem cells (SCs) of different linages, must provide the necessary  
number of cells for proper functioning, in order to get an adequate cell renewal and/or the reconstruction in case of damage 
(or development during embryogenesis) for the different tissues.

The strategies and rhythm of the mitosis must be adapted 
to the requirements for each specific case. For instance, during embryonic development, as a rule, the first mitosis must create a sufficient 
number of SC (through SC-SC symmetric divisions) in order to guarantee the high production of different cells at later times when the asymmetric 
divisions (SC -differentiated) in the intermediate stages take place. The symmetric (differentiated-differentiated) divisions in 
the final stage are those that prevail \cite{Taka,Barton}.

Cell renewal in homeostatic regimen is quite different. 
Here the strategy must lead to a stable (although fluctuating) number of different cell populations to compensate the disappearance of  
differentiated cells produced in their normal stage \cite{Klein}.

Colonic crypts are finger-shaped formations where the renewal processes take place. At the bottom of each crypt there is a small set of SCs 
that gives origin to a set of cells, known as transit amplifying cells (TACs), that perform between four and six divisions 
before the differentiated cells of the lineage are generated (secretory lineages: globet cells, enteroendocrine cells and absortive lineage: enterocytes). 
While mitosis develops, cells migrate from the bottom of the crypt up to their mouth, so that the already differentiated ones end up being lost at 
the intestinal lumen. In the small intestine, the SCs originate another type of secretory cell (Paneth cells) that migrates 
towards the bottom of the crypt. Paneth cells are absent or rare in the colon\cite{Ouelle} and their lifetime is quite higher than the other phenotypes of 
the lineage\cite{Alberts}. Wnt signaling in the crypts maintains the stem cells population, whereas Notch signaling leads to the diversification 
of stem cells progeny and controls the number that is committed to a secretory or absorptive fate \cite{Alberts,vanderFlier}.

Today techniques provide many observations in mice, so in the present work we focus on them. One striking aspect 
of the experimental results is the huge dispersion of data reported by different authors. For example: Ref.
\citen{Potten1} establish that the number of SCs per crypt is between 4 and 16; Ref. \citen{Paulus} reports between 4 and 8;  Ref. \citen{Li} reports 4; 
Ref. \citen{Yatabe} between 1 and 40 and Refs. \citen{Potten2}, \citen{Barker} between 4 and 6. The same pattern reappears concerning the total number 
of cells per crypt (Ref. \citen{Potten1}:  235-250; Ref. \citen{Drasdo}, 300).The fraction of proliferative cells by 
crypt is 2/3 in Ref. \citen{Potten1}, between 1/2 and 4/5 in Ref. \citen{Li}, 1/5 in Ref. \citen{Levin} and 1/3 in 
Refs. \citen{Ro} and \citen{Katz}. We could follow the list regarding the duration of the cell cycle of the SC, TAC and so on. 

Although this fact could be explained since the experimental techniques and the markers for the genes that identify the SCs are 
different, it seems to us that it implies that fluctuations in the population of the crypt in homeostatic regimen are important.

There is an interesting observation regarding the dynamics in the crypt. Strong evidence has been found indicating its monoclonal character:
that is, in a homeostatic regime, all the cells of the crypt are descendants of a single SC \cite{Snippert,Lopez-Garcia,Simons}. By using a crypt model with mechanical interactions \cite{Leeuwen,Pitt},the transit to monoclonality has been studied through computer simulations\cite{Mirams}.
In this context it is natural to describe this dynamics as a Moran process \cite{Fletcher}. Basically, this stochastic process consists of a random substitution produced by joint events of disappearance of a SC 
(through a TAC-TAC symmetric division) followed by the birth of another SC (SC-SC symmetric division) in a set of $N$ stem cells, where $N$ is fixed. Possible asymmetric SC-TAC divisions do not affect the clonal composition of the crypt. They only produce a delay in the transit  monoclonality. 

However, this type of model is incompatible with the assumption that $N$ could fluctuate 
in homeostatic conditions, since each stochastic event (disappearance-birth) preserves $N$. For example, two or more disappearances (births) in a row are 
not considered.

On the other hand, it has been speculated recently that the subdivision of the intestinal epithelium into millions of small niches, with few stem cells,
constitutes a natural barrier for the fixation of better adapted  mutations (that could be cancerous) due to the fact that  in small populations the 
neutral dynamics can prevail over the Darwinian adaptation \cite{Kang}. Therefore it is of interest to analyse the possible fixation of mutant clones 
in the framework of a neutral drift.

To study the dynamics of cell populations in the crypt (in homeostatic regime, cancerogenesis or damage), 
 two kinds of mathematical models have been developed: spatial models and compartment models \cite{vanLeeuwen,DeMatteis,Carulli}.

We propose a new compartment model for colonic crypts, taking into acount only the temporal evolution of the cell populations.
The model is a system of ordinary differential equations and the corresponding stationary solutions 
would correspond to the mean values of the cell populations. However, given the small number of cells involved, fluctuations will be important and thus we prefer to interpret these equations through their master equations. In this way we can study the stability of the stationary solutions and thus guarantee that the fluctuations would not lead, for example, to the extinction of the crypt.

In Sec.\ref{sec:model} we present the model for the colonic crypts. Section \ref{sec:first} is devoted to the first two compartments (SC and TAC of first generation).
In Sec. \ref{sec:Res} we show the results for those compartments and we study the neutral dynamics that leads to  monoclonality assuming a normal crypt 
(that is admitting equal fitness for all stem cells). We then assume the presence of different kinds of a mutant stem cell and we explore this effect.
Section \ref{sec:resto} refers to the other compartments. Finally in Sec. \ref{sec:end} we discuss the results and draw some conclusions.
In three Appendixes we present the algorithm to solve the master equation (\ref{sec:gille}), the nullclines of the differential equation (\ref{Null}) and
an example showing the relevance of studying the stability of compartment models in the presence of fluctuations or with the addition of intrinsic noise (\ref{fue}).

\section{\label{sec:model} The model}
The proposed model has eight compartments. One of them corresponds to SC, the next six to each generation of TACs (TAC$_{i}, i=1...6$) and the last one, for fully 
differentiated cells (DC).

The number of SCs ($S$) can be increased by means of SC-SC symmetric divisions with a rate of $M_{S,S}$ and by means of a cell reprogramming process (plasticity) 
of a TAC$_1$  where $P$ is the rate of such process. The processes that decrease $S$ are: the
symmetric divisions of SC giving two TAC$_1$ at a rate $M_{T_1, T_1}$ and the
apoptosis processes with a rate $A_S$. This leads to:

\begin{equation}
\frac{dS}{dt}=M_{S,S} S - M_{T_{1},T_{1}} S - A_{S} S + P T_{1} \, , \label{eq:00}
\end{equation}

The number of TAC$_1$s ($ T_1$) is increased by the TAC$_1$-TAC$_1$ symmetric divisions 
of SCs as discussed above and by the asymmetric divisions of SCs that produce a SC and a
TAC$_1$ at rate $M_{S,T_1}$. On the other hand, $T_1$
may decrease due to  TAC$_2$-TAC$_2$ symmetric divisions of TAC$_1$s at rate of 
$L_{T_1}$ and the apoptosis processes at rate $A_{T_1}$. All this leads to equation:

\begin{equation}
\frac{dT_{1}}{dt}=2 M_{T_{1},T_{1}} S + M_{S,T_{1}} S - L_{T_1} T_{1} - A_{T_{1}} T_{1} \, . \label{eq:01}
\end{equation}

These two compartments ($S$ and $T_1$) are the fundamental ones and can be considered the "engine" of the crypt.
The TAC$_2$s and TAC$_3$s  are increased at a rate equal to twice the symmetric divisions of the previous generation and,
their number decreases when they divide symmetrically into two cells of the subsequent generation.This is displayed as:

\begin{equation}
\frac{dT_{2}}{dt}=2 L_{T_{1}} T_{1} - L_{T_{2}} T_{2} \, , \label{eq:02}
\end{equation}

\begin{equation}
\frac{dT_{3}}{dt}=2 L_{T_{3}} T_{2} - L_{T_{3}} T_{3} \, . \label{eq:03}
\end{equation}

For the fourth generation, in addition to the symmetrical divisions of TAC$_4$ (TAC$_5$-TAC$_5$) that feed the population of the fifth generation
, the probability of complete differentiation must be considered through DC-DC symmetric divisions or DC-TAC$_5$ asymmetric divisions (these do not change the number of TAC$_4$s but they change the number of TAC$_5$s). 
This is taken into account by the factors $\alpha_{T_5,T_5}, \alpha_{D,D}$  and $\alpha_{T_5,D}$ which are the fraction of each division mentioned above.

The processes involved in the balance for TAC$_5$s are similar to those of TAC$_4$s. In this case
$\alpha^\prime_{T_6,T_6}, \alpha^\prime_{D,D}$ and $\alpha^\prime_{T_6,D}$
are the fraction of each type of division. The last generation, TAC$_6$s, only produces DCs. Therefore this can be written as:

\begin{equation}
\frac{dT_{4}}{dt}=2 L_{T_{3}} T_{3} - (\alpha_{T_5,T_5}+\alpha_{D,D}) L_{T_{4}} T_{4} \, , \label{eq:04}
\end{equation}

\begin{equation}
\frac{dT_{5}}{dt}=(2 \alpha_{T_5,T_5}+ \alpha_{T_5,D}) L_{T_{4}} T_{4} - (\alpha^{\prime}_{T_6,T_6}+\alpha^{\prime}_{D,D}) \, ,
L_{T_{5}} T_{5}, \label{eq:05}
\end{equation}

\begin{equation}
\frac{dT_{6}}{dt}=(2 \alpha^{\prime}_{T_6,T_6}+ \alpha^{\prime}_{T_6,D}) L_{T_{5}} T_{5}- L_{T_6} T_{6} \, . \label{eq:06}
\end{equation}
Finally, the DC compartment is fed by the divisions of TAC$_4$s, TAC$ _5$s and TAC$_6$s, and $A_D$ accounts for apoptosis or entry into the intestinal lumen.

\begin{equation}
\frac{dD}{dt}=2 L_{6} T_{6}+(2 \alpha_{D,D}+\alpha_{T_5,D}) L_{T_4} T_{4}+ (2 \alpha^{\prime}_{D,D}+\alpha^{\prime}_{T_6,D}) \, .
 L_{T_{5}} T_{5} - A_{D} D. \label{eq:07}
\end{equation}

\begin{figure*}[!t]
\centerline{\includegraphics[scale=0.9,clip=true,angle=0]{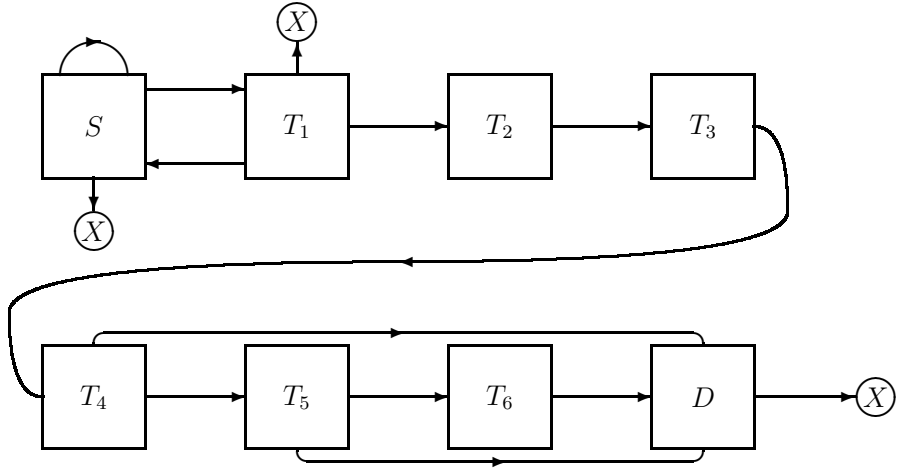}}
\caption{Sketch of the model.}
\label{fig:01}
\end{figure*}

A sketch of the model is shown in Fig. (\ref{fig:01}).

Note that it is necessary to consider each generation of TACs in a different compartment. The division of each mother TAC  produces cells that are diferent to it (their daughters are more commited in the differentiation process). If the divisions of TACs would feed into the same compartment of those TACs, we would be introducing immortal TACs, which contradicts its transient nature.

The system of Eqs. \ref{eq:00}-\ref{eq:07} should lead, in the stationary regime, to a finite population for each kind of cells. It is worth mentioning that the coefficients representing the 
rates for each process cannot be constant. If so, we would have a system of linear first order differential equations, whose general solution involves 
exponentially increasing populations over time. These could be eliminated with an appropriate choice of parameters. Even then, our model should be 
completely unsatisfactory since any fluctuation of these parameters around those values (quite small indeed) would lead to the exponential growth of some of the 
populations. The solution would be structurally unstable and therefore not acceptable as a description of a biological system in a homeostatic regime \cite{Johnston}.
Furthermore, the solution should also be stable against internal noise present in the system (see \ref{fue}). 

The number of cells that emerge from the stationary solutions of the equations are continuous, but we are dealing with discrete quantities, therefore they must be interpreted as an 
average value of its temporal evolution. This behaviour may be far away from the real one if the populations involved are small, as in the crypts (see \ref{fue}).

Equations (\ref{eq:00}) and (\ref{eq:01}), that correspond to the SC and TAC$_1$ compartments, are strongly coupled and they feed the rest of the compartments.
Thus we will establish the parameters and the solutions for these two equations and, then, we will let evolve the solution with the remaining six.

\section{\label{sec:first} The SC and TAC$_1$ compartments.}

In this section we will consider that the processes involved in the model are regulated by a population-dependent feedback. As it was already mentioned, such feedback 
should not only guarantee the stability of the solution against the parameters fluctuations but also, against intrinsic noise of the equations. This regulation is included 
through Hill functions as:

\begin{equation}
M_{S,S}(S,T_1)=\frac{K_{S,S}{S_o}^{n}}{({S_o}^n+{S}^n)} \frac{T_{1}^n}{({T_{o1}}^n+{T_{1}}^n)} \, \,, \label{eq:08}
\end{equation}
\begin{equation}
M_{T_{1},T_{1}}(S,T_1)=\frac{K_{T_{1},T_{1}} {S}^{n}}{({S_o}^n+{S}^n)} \frac{T_{o1}^n}{({T_o1}^n+{T_{1}}^n)}    \, \, , \label{eq:09}
\end{equation}
\begin{equation}
A_{S}(S)=\frac{\lambda {S}^{n}}{({S_o}^n+{S}^n)} \, \, , \label{eq:10}
\end{equation}
\begin{equation}
P(S)=\frac{\beta {S_o}^{n}}{({S_o}^n+{S}^n)} \, \, , \label{eq:11}
\end{equation}
\begin{equation}
M_{S,T_{1}}= K_{S,T_{1}} \, , \label{eq:12}
\end{equation}
\begin{equation}
L_{T_1}(T_1)=\frac{\beta^{\prime}{T_{1}}^{n}}{({T_{o1}}^n+{T_{1}}^n)} \, \, , \label{eq:13}
\end{equation}
\begin{equation}
A_{T_{1}}(T_1)=\frac{\lambda^{\prime} {T_{1}}^{n}}{({T_{o1}}^n+{T_{1}}^n)} \, \, . \label{eq:14}
\end{equation}

Here $K_{S,S}, K_{T_{1},T_{1}},K_{S,T_{1}},S_o,T_{o1},\lambda,\lambda^{\prime},\beta,n$ are parameters to be determined. Let us remark that the rate of SC-TAC$_1$ 
asymmetric division is taken as a constant (i.e. not regulated or much less regulated than the other considered processes), since this process does not change the number of SC. 
We considered the same exponent $n$ for all Hill functions and it will be determined as the smaller integer that leads to a stable solution against the internal noise.
 
The other parameters will be chosen in such a way that the number of SC ($S_s$) and TAC$_1$ ($T_{s1}$) for the stationary solution are close to 15 for each case. 
The rate of cell division for the SC for the stationary solution results $ 1 / 24$h . That is:
\begin{equation}
M_{S,S}(S_s,T_{s1})+M_{T_{1},T_{1}}(S_s,T_{s1})+K_{S,T_{1}} \approx \frac{1}{24 \mbox{h}} \, \, . \label{eq:15}
\end{equation}

The rate for the cellular divisions of TAC$_1$ is $1/19$ h  for the same stage, that is:

\begin{equation}
L_{T_{1}}(T_{s1})+P(S_s) \approx \frac{1}{19 \mbox{h} } \, \, .  \label{eq:16}
\end{equation}

The cell cycle length of $TAC_{1}$ is much higher than those reported in the literature (10h-12h). It is not clear that 
these determinations refer to $TAC_{1}$. 
We assume that it is an average of the cell cycle length of the 6 generations of TACs. Later, in Sec.\ref{sec:resto} when 
the parameters are defined for the other compartments, we will adjust them so that the average of the cycle length of the TACs results 10h. 

A fundamental issue considered here is the speed of evolution towards monoclonality at short times. In our model, this 
is quite sensitive to parameter $K_{S,T_{1}}$ (the rate of asymmetric divisions). In fact, if all the divisions 
were of this kind, there would be no transition to the monoclonality while if were not such divisions, this 
transition would be very fast.
This speed was measured experimentally and for 15 initial marked SCs (15 initial clones
from one SC cell each) the average number of SCs in each surviving clone is approximately 6 cells at 14 days \cite{Snippert}.
$ K_{S,T1}$ must be equal to zero to reproduce this result.
In conclusion, in this model, the process of renovation of the crypts in the homeostatic regimen, is the consequence of a regulated balance between the 
symmetric divisions of SCs of both types (SC-SC and TAC $ _1 $ -TAC $ _1 $). It is worth noting that this does not require the divisions that lead to the disappearance of one SC (TAC $ _1 $ -TAC $ _1 $), are followed by a division leading to a new SC (SC-SC). Indeed if the differential equations are considered as master equations for stochastic processes, several divisions of the same type may occur causing large fluctuations in the involved populations.

The parameters of the Eqs.(\ref{eq:00}) and (\ref{eq:01}) necessary to obtain all the results 
shown below are given in table (\ref{Tab:01}).

\begin{table}
\caption{Parameters used in the calculations for Eqs.(\ref{eq:00}) and (\ref{eq:01})}

\begin{center}
{\begin{tabular}{|c|c|c|c|c|} \hline  

$K_{S,S}$ (h$^{-1}$)& $K_{T_{1},T_{1}}$(h$^{-1}$) &$K_{S,T_{1}}$(h$^{-1}$)& $S_o$ &$T_{o1}$  \\

\hline \hline
$7.53\times10^{-2}$&$9.58\times10^{-2}$ & 0 & $17.56$ & 19\\
\hline

\end{tabular}}{}

\vspace{0.3cm}

{\begin{tabular}{|c|c|c|c|c|} \hline 

$\lambda$ (h$^{-1}$) &$\lambda^{\prime}$(h$^{-1}$)&$\beta$(h$^{-1}$)& $\beta^{\prime}$(h$^{-1}$)& $n$ \\

\hline \hline
$1.33\times10^{-4}$ &$1.44\times10^{-4}$&$1.1\times10^{-2}$  & $1.15\times10^{-1}$ & 2\\
\hline

\end{tabular}}{}

\end{center}

\label{Tab:01} 
\end{table}

\section{\label{sec:Res} Results}

\subsection{ \label{subsec:global} Time evolution of populations in homeostatic regime.}

In this subsection we show the results that emerge from the equations of the populations of SCs and TAC$_1$s. The solution for 
the differential equations has a stable fixed point $(S_{s}=15, T_{s1}=15.62)$ (see \ref{subsec:normal} ) and it describes the behavior of 
the average population in homeostatic conditions. The solution corresponding to master equations that emerge from these differential equations, 
is solved using an algorithm based on Gillespie's method \cite{Gilli} (see \ref{sec:gille}).

\begin{figure*}[!t]
\centerline{\includegraphics[scale=0.5,clip=true,angle=0]{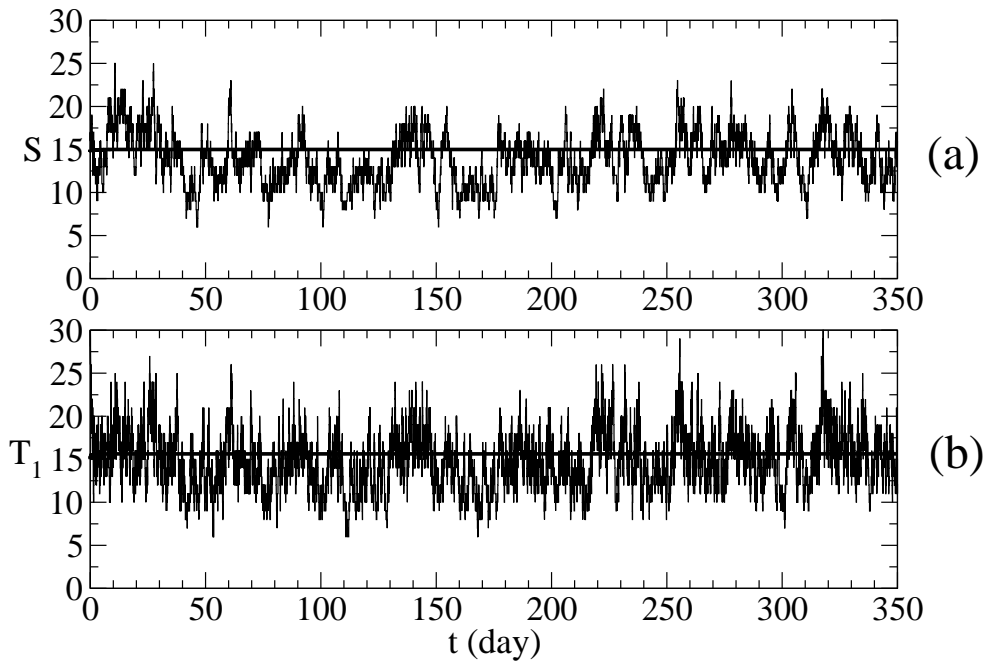}}
\caption{(a) Time evolution of SC population. The horizontal line in 15 corresponds to the stationary solution of the
 differential equations. The noisy curve around this solution corresponds to one realization obtained by considering the
differential equations as a master equation.
(b) The same as in (a) but referring to the TAC$_1$ population}

\label{fig:02}
\end{figure*}

Figure (\ref{fig:02}) (a) and (b) shows the solution of equations (Eqs.(\ref{eq:00}) and (\ref{eq:01})) for the populations of 
SCs and TAC$_1$s as a function of time. They are displayed as thick lines parallel to the $t$ axis in each graph. Also it is 
shown a realization of the solutions obtained by using the algorithm mentioned above.
It can be observed that although the populations are highly fluctuating in time, such fluctuations do not lead to the extinction 
of the crypt. We have tested it on $10^6$ different realization.

\begin{figure*}[!t]
\centerline{\includegraphics[scale=0.5,clip=true,angle=0]{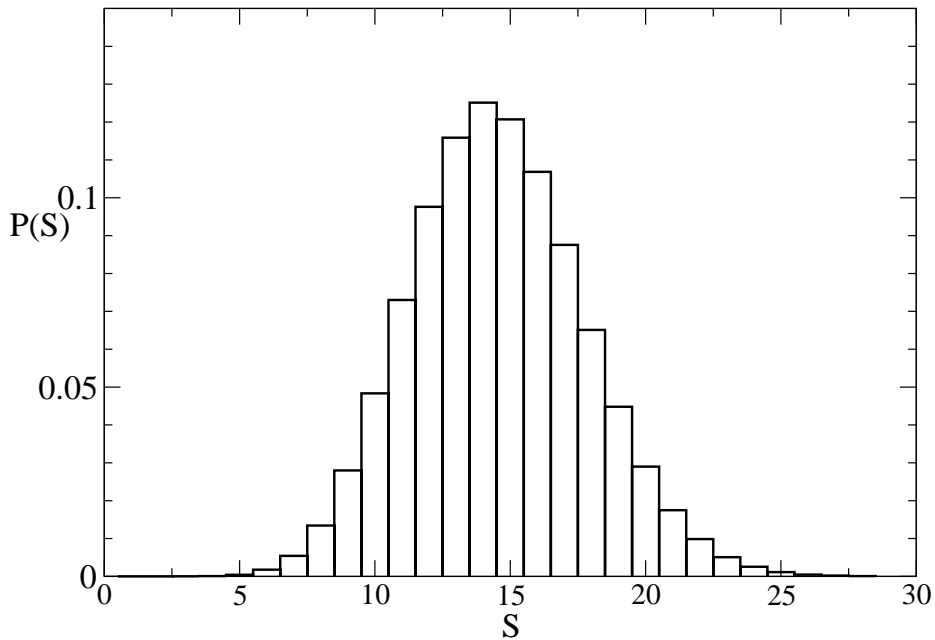}}
\caption{Probability distribution of SC population in homeostatic regime obtained from frequencies over 10$^6$ cripts.}
\label{fig:03}
\end{figure*}

To characterize these fluctuations, the probability distribution $P(S)$ for $S$ was calculated from the 
frequencies coming from these $10^6$ realizations. The distribution is shown in the Fig.(\ref{fig:03}).
Its  mean value, $\langle S \rangle$, and the standard deviation, $s$, is given by:

\begin{equation}
\langle S \rangle= \sum_{S} S P(S) =  14.54 \, , 
\end{equation}

\begin{equation}
s= \sqrt{ \big \langle (S-\langle S \rangle)^{2} \big \rangle}  =  3.2  \, . 
\end{equation}

\subsection{ \label{subsec:neutral} Neutral drift and transit to monoclonality}
We introduce the conditional probability that cell $j$ participates in a process given that the 
process has occurred.
As can be observed, the total number of stem cells $S(t)$ is highly fluctuating.
Assuming that SCs are identical (equal fitness) for the processes considered in the model the conditional probabilities for symmetric division (TAC$ _1$-TAC$_1$ or SC-SC), apoptosis process etc. will be:

\begin{eqnarray}
P_{S}(j / \mbox{(SC-SC)}) & = & \frac{1}{S (t)} \, ,  \nonumber\\
P_{S}(j / \mbox{(SC-TAC$_1$)}) & = & \frac{1}{S (t)} \, ,  \label{eq:17} \\
P_{S}(j / \mbox{(Apop.)}) & = & \frac{1}{S (t)} \, ; \nonumber
\end{eqnarray}

and the same for TAC$_1$:

\begin{eqnarray}
P_{T_1}(j / \mbox{(Plast.)}) & = & \frac{1}{T_{1}(t)} \, \, ,  \nonumber\\
P_{T_1}(j / \mbox{(TAC$_2$-TAC$_2$)}) & = & \frac{1}{T_{1}(t)} \, \, , \label{eq:18} \\
P_{T_1}(j / \mbox{(Apop.)}) & = & \frac{1}{T_{1}(t)} \, \, . \nonumber
\end{eqnarray}

If at $t=0$ there are $S(t=0)=N$ with equal fitness, each of these cells gives
rise to a clone. In particular, at $t=0$ there will be $T_{1}(t=0)=N^{\prime}$ originated by SCs.
We must determine how the population of their offspring
evolves as time evolves. At time $t$, the total populations $S(t)$ and $T_{1}(t)$ can be written as:
\begin{eqnarray}
S(t) & = & \sum_{i} s_{i}(t) \,,  \nonumber \\
T_{1}(t) & = & \sum_{i} t_{i}(t) \, ;
\end{eqnarray}
where $s_{i}(t)$ ($t_{i}(t)$) are the SCs (TAC$_1$s) present at time $t$ which are descendent of each original SC. That is 
the i-th clone 
has $s_{i}(t)$ SCs and $t_{i}(t)$ TAC$_1$s at time $t$. We can extend the above conditional probabilities for each clone in the involved processes.

\begin{eqnarray}
P_{S}(i \mbox{-th clone}/ \mbox{(SC-SC)}) & = & \frac{s_{i}(t)}{S (t)} \, \, , \nonumber\\
P_{S}(i \mbox{-th clone} / \mbox{(TAC$_1$-TAC$_1$)}) & = & \frac{s_{i}(t) }{S (t)} \, \, ,  \nonumber\\
P_{S}(i \mbox{-th clone} / \mbox{(Apop.)}) & = & \frac{s_{i}(t) }{S (t)} \, \, , \label{eq:19}  \\
P_{T_1}(i \mbox{-th clone} / \mbox{(Plast.)}) & = & \frac{t_{i}(t)  }{T_{1}(t)} \, \, ,  \nonumber\\
P_{T_1}(i \mbox{-th clone} / \mbox{(TAC$_2$-TAC$_2$)}) & = & \frac{t_{i}(t) }{T_{1}(t)} \, \ , \nonumber\\
P_{T_1}(i \mbox{-th clone} / \mbox{(Apop.)}) & = & \frac{t_{i}(t)}{T_{1}(t)} \, \, . \nonumber
\end{eqnarray}

Thus, when the equations are solved by using the stochastic algorithm, in addition to toss the process that occurs at 
each step, we choose at random the clone that participates in each case.

We characterize the dynamics calculating two quantities as a function of time. 

The average size of clones in the $SC$ compartment, ($q$), is defined by:
\begin{equation}
q(t) = \frac{\sum_{i} s_{i}(t)}{M(t)} \, \, ;
\end{equation}

\begin{figure*}[!t]
\centerline{\includegraphics[scale=0.5,clip=true,angle=0]{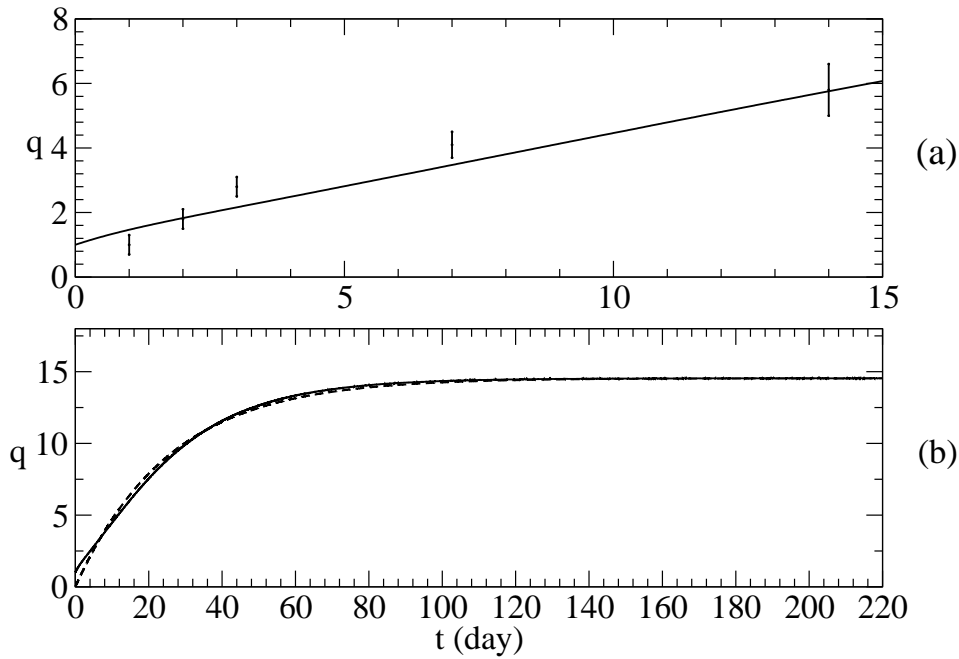}}
\caption{(a) Mean average size $\langle q \rangle$ as a function of time for short times. 
The points correspond to experimental data taken from \cite{Snippert}.
(b) Same curve as in (a) but until times in which the homeostatic regimen has been reached (continuous curve). 
The dashed line curve corresponds to the best fit of the functional form 
 $14.54 (1-\exp(-t/\tau))$ ($\tau=25.56$ day).}
 
\label{fig:04}
\end{figure*}
where $M(t)$ is the number of clones at times $t$. Initially we set $q=1$ since we have $N$ clones with one SC each. 
Due to fluctuations, the average size $q$ for a single crypt is equally fluctuating. 
Therefore, we calculate its mean behavior over $10^6$ crypts $\langle q(t) \rangle $.
We started with 15 SCs at $t=0$. Figure (\ref{fig:04}) shows the time evolution of $\langle q \rangle $. 
The panel (a) shows $\langle q \rangle$ for short times together experimental data taken from Ref. \citen{Snippert}.
Let us remark that the 
inclusion of SC-TAC$_1$ asymmetric division (i.e $K_{S,T_1} > 0$) in Eq.(\ref{eq:02}) decrease the slope of the curve. 
The panel (b) shows the complete curve.

To evaluate the characteristic time of transit to monoclonality, we have fitted the curve as:

\begin{equation}
\langle q(t) \rangle = 14.54 (1-\exp(-t/\tau)) \, , 
\end{equation} 

being $\tau=25.56$ day.

\begin{figure*}[!t]
\centerline{\includegraphics[scale=0.5,clip=true,angle=0]{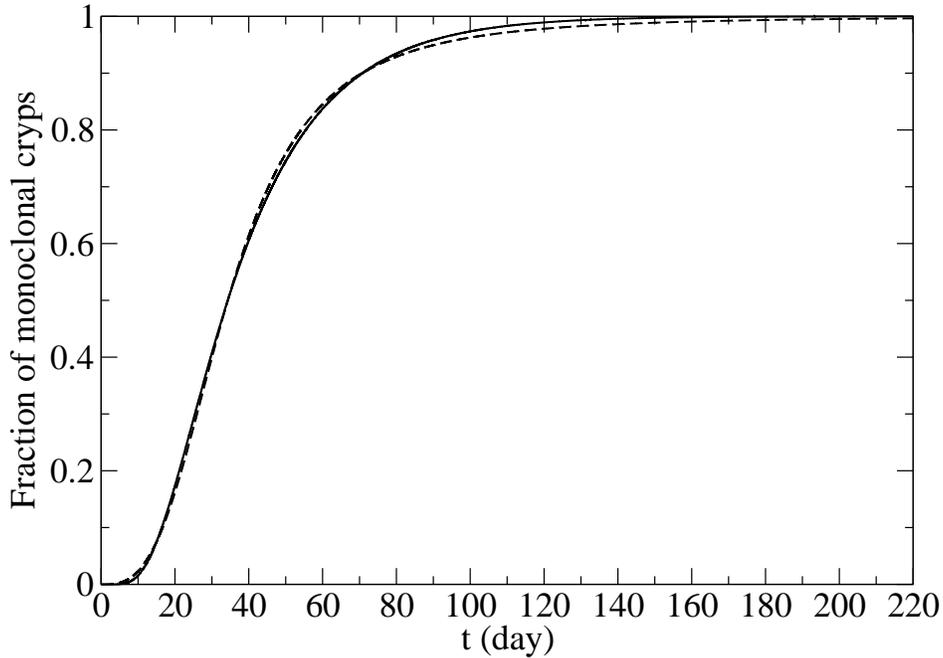}}
\caption{Fraction of monoclonal crypts as a function of time (continuous line curve).To determine a characteristic 
time, we have fitted the curve with a Hill function ($\frac{t^p}{(t_{m}^p+t^{p})}$) with 
$t_{m}=34.32$ day and $p=3.04$ (dashed line curve).}
\label{fig:05}
\end{figure*}

Another way to evaluate the transit time towards the monoclonality is by 
means of the calculation of the fraction of monoclonal crypts $f$ as a function of time $t$ from an ensamble of them.
Figure (\ref{fig:05}) shows such fraction. To calculate the characteristic time, we have fitted the curve as:
\begin{equation}
f(t) = \frac{t^p}{(t_{m}^p+t^{p})} \, ; 
\end{equation}

and the best fit leads to $p=3.04$ and the time when half of crypts will be monoclonal results $t_{m}=34.32$ day.

\subsection{ \label{subsec:muta} Neutral drift and mutations}

The neutral drift leads to monoclonal crypts. In the results shown in the previous subsection, the initial state
of the crypts had 15 distinguishable SCs and 15 TAC$_1$ (fifteen clones of one SC and one TAC$_1$ each). After evolution the final state
was monoclonal comprising all the SCs present (and all the TAC$_1$). As the initial 15 clones are equally adapted, the surviving clone results 
from a symmetry breaking process induced by stochastic fluctuations. The relative monoclonality 
frequencies of each clone $ f (i), i = 1,2, ... ,15 $ are equal, in particular $ f (i) = 1/15 $, as can be seen in Fig.(\ref{fig:06})(a).

Let us now consider that due to some mutation, some of the initial 15 clones has different behavior in the
processes. 
Assuming, without loss of generality, the clone is labeled as 15. The conditional probabilities given by 
Eqs.(\ref{eq:19}) are no longer valid, therefore we  modify them as follows:
\begin{eqnarray}
P_{S}(15 \mbox{\,clone} / \mbox{(SC-SC)}) & = & \frac{ m_1 s_{15}(t)}{\sum_{j=1}^{14} s_{j}(t)+m_1 s_{15}(t)} \, \, ,  \nonumber\\
P_{S}(15 \mbox{\,clone} / \mbox{(TAC$_1$-TAC$_1$)}) & = & \frac{m_2 s_{15}(t) }{\sum_{j=1}^{14} s_{j}(t)+m_2 s_{15}(t)} \, \, ,      \nonumber\\
P_{S}(15 \mbox{\,clone} / \mbox{(Apop.)}) & = & \frac{m_3 s_{15}(t) }{\sum_{j=1}^{14} s_{j}(t)+m_3 s_{15}(t)} \, \, , \label{eq:20}  \\
P_{T_1}(15 \mbox{\,clone} / \mbox{(Plast.)}) & = & \frac{m^{\prime}_1 t_{15}(t)  }{\sum_{j=1}^{14} t_{j}(t)+m^{\prime}_1 t_{15}(t)} \, \, ,  \nonumber\\
P_{T_1}(15 \mbox{\,clone} / \mbox{(TAC$_2$-TAC$_2$)}) & = & \frac{m^{\prime}_2 t_{15}(t) }{\sum_{j=1}^{14} t_{j}(t)+m^{\prime}_2 t_{15}(t)} \, \, , \nonumber\\
P_{T_1}(15 \mbox{\,clone} / \mbox{(Apop.)}) & = & \frac{m^{\prime}_3 t_{15}(t)}{\sum_{j=1}^{14} t_{j}(t)+m^{\prime}_3 t_{15}(t)}\, \, ; \nonumber
\end{eqnarray}
where $m_{i},m^{\prime}_{i}, i=1,2,3$ are parameters related to the mutation. 
For the other clones ($i=1,...,14$)  the modifications are:
\begin{eqnarray}
P_{S}(i \mbox{-th clone} / \mbox{(SC-SC)}) & = & \frac{s_{i}(t)}{\sum_{j=1}^{14} s_{j}(t)+m_1 s_{15}(t)} \, \, ,  \nonumber\\
P_{S}(i \mbox{-th clone} / \mbox{(TAC$_1$-TAC$_1$)}) & = & \frac{s_{i}(t) }{\sum_{j=1}^{14} 
s_{j}(t)+m_2 s_{15}(t)} \, \, ,      \nonumber\\
P_{S}(i \mbox{-th clone} / \mbox{(Apop.)}) & = & \frac{s_{i}(t) }{\sum_{j=1}^{14} s_{j}(t)+m_3 s_{15}(t)} \label{eq:21} \, \, ,  \\
P_{T_1}(i \mbox{-th clone} / \mbox{(Plast.)}) & = & \frac{t_{i}(t)  }{\sum_{j=1}^{14} t_{j}(t)+m^{\prime}_1 t_{15}(t)} \, \, ,  \nonumber\\
P_{T_1}(i \mbox{-th clone} / \mbox{(TAC$_2$-TAC$_2$)}) & = & \frac{t_{i}(t) }{\sum_{j=1}^{14} 
t_{j}(t)+m^{\prime}_2 t_{15}(t)} \, \, , \nonumber\\
P_{T_1}(i \mbox{-th clone} / \mbox{(Apop.)}) & = & \frac{t_{i}(t)}{\sum_{j=1}^{14} t_{j}(t)+m^{\prime}_3 t_{15}(t)} \, \, . \nonumber
\end{eqnarray}
Let us remark that if we set all $m_i$ and $m^\prime_i$ equal to one in Eqs.(\ref{eq:20}) and (\ref{eq:21}), the Eqs.(\ref{eq:19}) are 
recovered.

Starting from $s_i=1, t_i=1, (i=1,2,...,15)$ and as a result of the neutral drift for large time $t$, that guarantees the establishment of 
homeostatic regime, we obtain $s_{i}(t)=0, t_{i}(t)=0, \forall i\neq k$ and $S(t)=s_{k}(t), T_{1}(t)=t_{k}(t)$. If $k \neq 15$, the crypt will 
be normal, otherwise, the crypt becomes mutant. In this case, the differential equations (and the master equations) describing their evolution 
will be the same but with new parameters given by:

\begin{figure*}[!t]
\centerline{\includegraphics[scale=0.5,clip=true,angle=0]{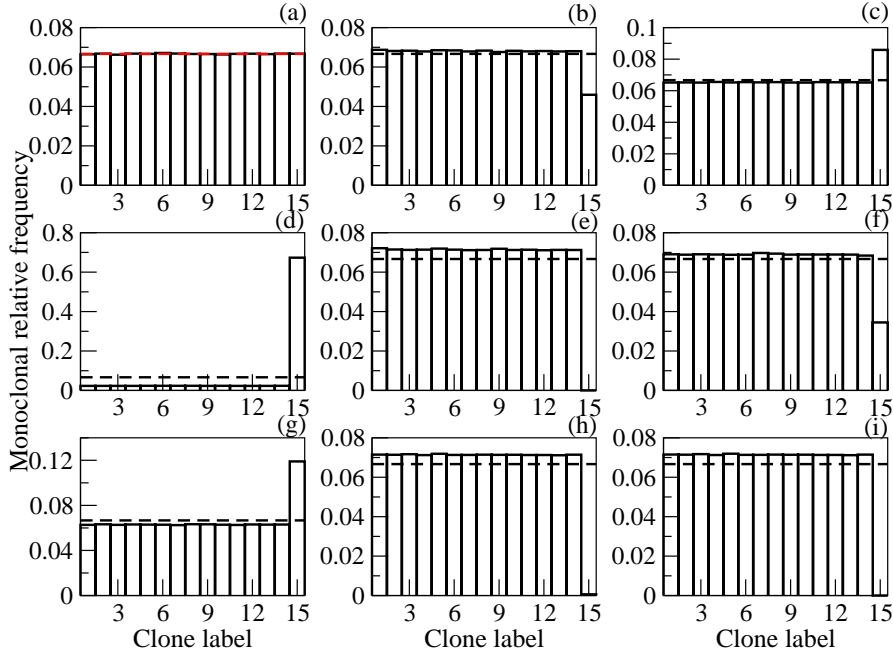}}
\caption{Relative fixation frequency of each clone. (a) All clones are identical.
(b) Clone 15 is a mutant with $m_i=m^\prime_i=4, i = 1,2,3$. (c) Clone 15 is a mutant with $m_i=m^{\prime}_{i}=0.25, i = 1,2,3$. 
(d) Clone 15 is a mutant with $m_1=1.1,m_2=0.9, m_3=m^{\prime}_{i}, i = 1,2,3$. (e) Clone 15 is a mutant $m_i=m^{\prime}_{3}=1, i = 1,2,3, m^{\prime}_{1}=1.1,
m^\prime(2)= 0.9$.(f) Clone 15 is a mutant with $m_{i}=m^{\prime}_{3}=1, i = 1,2,3, m^{\prime}_{1} = 0.9, m^{\prime}_{2}=1.1$. 
(g) Clone 15 is a mutant with $m_{i}=m^{\prime}_{3}=1, i = 1,2,3, m^{\prime}_{1} = 1.1, m^{\prime}_{2}=0.9$.
(h) Clone 15 is a mutant with $m_{i}=m^{\prime}_{2}=1m^{\prime}_{3}=1, i = 1,2,3, m^{\prime}_{1} =0$ (see Fig.\ref{fig:07}(b)).
(i) The same as in (h) but for surviving crypts (see Fig.\ref{fig:07}(c).}
\label{fig:06}
\end{figure*}

\begin{eqnarray}
K^{(m)}_{S,S}=m_{1} K_{S,S}, &  K^{(m)}_{T_1,T_1}=m_{2} K_{T_1,T_1}, & \lambda^{(m)}=m_{3} \lambda, \\
\beta^{(m)}=m^{\prime}_{1} \beta, & \beta^{\prime (m)}=m^{\prime}_{1} \beta^{\prime}, & \lambda^{\prime (m)}=m^{\prime}_{3} \lambda^{\prime}.
\end{eqnarray}

This means that the fixation of a mutant clone could produce the modification of the equilibrium populations, the possible disappearance of the equilibrium or the 
extinction of the crypt.

In the following we will consider seven possible mutations.

\begin{figure*}[!t]
\centerline{\includegraphics[scale=0.5,clip=true,angle=0]{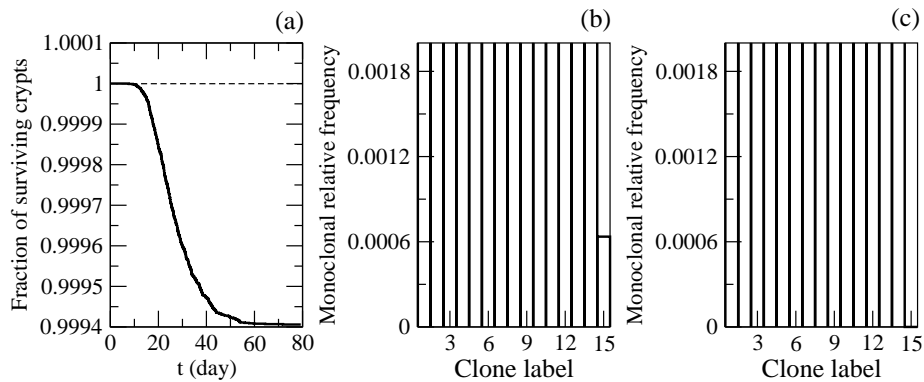}}
\caption{(a) Fraction of surviving crypts as a function of time when the mutant clone has no plasticity ($m^{\prime}_{1}=0$). 
The crypts that are extinguished correspond to the cases in which the mutant clone has itself imposed.
(b) Graph equal to that of Fig.\ref{fig:06}(h) but with a more refined scale for the abscissa.
(c) Graph equal to that of Fig.\ref{fig:06}(i) but with a more refined scale for the abscissa.}
\label{fig:07}
\end{figure*}

\begin{enumerate}
\item  {\boldmath {$m_{i}=m^{\prime}_{i} = 4, i=1,2,3$}}
\\
This means that the SCs and TAC$_1$s of clone 15 are four times as ``efficient" as the other clones in all processes considered. Clearly this 
mutation conspires against the fixation of the mutated clone, as can be seen in Fig.(\ref{fig:06})(b). 
The fixation of mutant clone does not change the equilibrium configuration but the crypt renewal is faster.

\item {\boldmath {$m_{i}=m^{\prime}_{i}=0.25, i=1,2,3$}}
\\ 
In this case, the mutated clone has a quarter of the "efficiency" of the others for all processes.
This mutation favors fixation of the clone (Fig.(\ref{fig:06})(c)). As above, this one does not change the equilibrium configuration but
the crypt renewal is slower.

\item {\boldmath {$m_{1}=1.5, m_{2}=0.5, m_{3}=m^{\prime}_{i}=1, i=1,2,3$}}
\\
Mutation increases the fraction of SC-SC mitosis and decreases the fraction of TAC$_1$-TAC$_1$ mitosis compared to the 
other clones. The monoclonal frequency of the mutated clone increases significantly (Fig.(\ref{fig:06})(d), 
notice the scale change on the abscissa). When the monoclonal regimen is established for the mutant, the fixed point of 
the differential equations results $(S_s=26.98,T_{s1}=17.87)$. Cell populations then fluctuate around these values.

\item {\boldmath {$m_1=0.5, m_2=1.5, m_3=m^{\prime}_{i} = 1, i=1,2,3$}}
\\
Mutation decreases the fraction of SC-SC mitosis and increases the fraction of TAC$_1$-TAC$_1$ mitosis compared to the 
other clones. The monoclonal frequency of the mutated clone decreases considerably ($2.2 \times 10^{-5}$). It is three orders of magnitude smaller than that of normal clones 
($7.2 \times 10^{-2}$) (Fig.(\ref{fig:06})(e). 
When the monoclonal regimen is established for the mutant, the fixed point of the differential equations results $(S_s=8.4,T_{s1}=11.22)$. 
Cell populations then fluctuate around these values.

\item {\boldmath {$m_{i}=m^{\prime}_{3}=1, i=1,2,3, m^{\prime}_{1}=0.9, m^{\prime}_{2}=1.1$}}
\\
In this case mutation changes the rates of TAC$_1$ rather than those of SC.
The mutated TAC$_1$s of the clone 15 have less plasticity and more TAC$_2$-TAC$_2$ mitosis than the others.
The monoclonal frequency of mutated clone decreases (Fig.(\ref{fig:06})(f)) and the fixed point is $S_s=13.5, T_{s1}=14$.

\item {\boldmath $m_i=m^{\prime}_3=1, i=1,2,3, m^{\prime}_1=1.1, m^{\prime}_2=0.9$}
\\
As in previous mutation, the rates of TAC$_1$s are modified, but not of the SC.
The mutated TAC$_1$s have more plasticity and less TAC$_2$-TAC$_2$ mitosis than the others.
The monoclonal frequency increases (Fig.(\ref{fig:06})(g)) and the fixed point is $S_s=16.67, T_{s1}=17.42$

\item {\boldmath $m_{i}=m^{\prime}_2=m^{\prime}_3=1, i=1,2,3, m^{\prime}_{1}=0$}
\\
In this case the mutant loses the plasticity completely. As can be seen in Fig. \ref{fig:06}(h), this produces a drastic drop in the probability of fixation of the mutant 
clone ($6.36 \times 10^{-4}$, two orders of magnitude smaller than that of normal clones: $7.14 \times 10^{-2}$, see Fig. \ref{fig:06}(h)). 
In addition, unlike the previous mutations, once the mutant monoclonality is established, the crypt becomes unstable and it is 
extinguished shortly thereafter (see Fig. \ref{fig:06}(i), that corresponds to the frequencies of surviving clones). To clarify this point 
in Fig.\ref{fig:07} (b) and (c) we reproduce the same graphs enlarging the scale of the abcissa. Figure \ref{fig:07}(a) shows the surviving crypt as a function 
of time. We can see that $6 \times 10^{-4}$ of the crypts are extinguished. This is the frequency of fixation for the mutant clone (as it is shown in Fig.\ref{fig:07} (b)). 
This behavior is discussed in \ref{subsec:mutant} with another point of view.

\end{enumerate}

\section{\label{sec:resto} The other compartments of the crypt}

In this section we will show the results corresponding to the rest of the compartments of the crypt 
(Eqs.\ref{eq:03}-\ref{eq:07}) which are feeding by the first two compartments.

One of the characteristics of TACs, unlike SCs, is that their offspring may not be identical to them since each 
division produces increasingly commited cells in the process of differentiation. Moreover, starting from the second generation of TACs, there is no processes of cell reprogramming (plasticity) that revert their fate.

\begin{figure*}[!t]
\centerline{\includegraphics[scale=0.6,clip=true,angle=0]{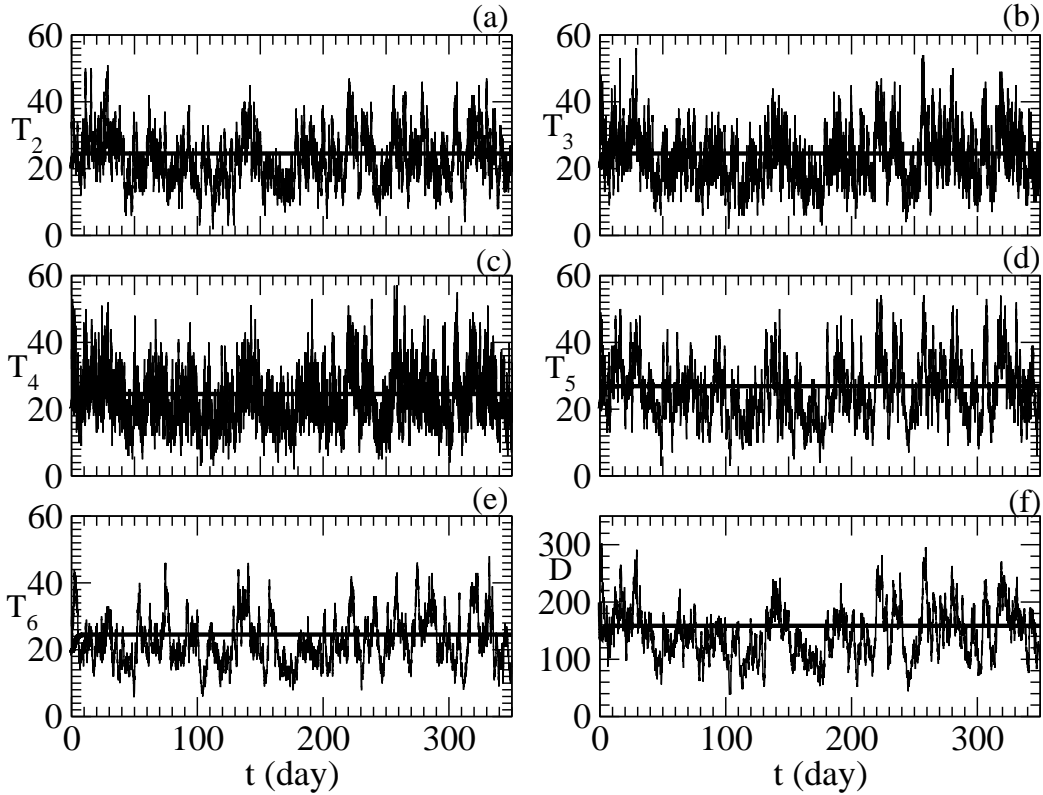}}
\caption{Cell populations as a function of time for:
(a) TAC$_2$. (b) TAC$_3$, (c) TAC$_4$, (d) TAC$_5$ (e) TAC$_6$ (f) Fully differentiated cells. The horizontal lines correspond to the equilibrium populations
according with the differential equations. The fluctuating lines correspond to a realization of the solution of the master equations.}
\label{fig:08}
\end{figure*}

That is why the processes that describe the equations can be compared to a slide in which enter TAC$_1$s and leave DCs. 
At the same time, complex processes occur through which, as a result of interactions between cells, external 
gradients, asymmetric inheritances and so on, it decides the final fate of each cell, giving the different phenotypes 
of DC that constitute the lineage.

\begin{table}
\caption{Parameters for Eqs.(\ref{eq:02})-(\ref{eq:07}) and (\ref{eq:22})}

\begin{center}
{\begin{tabular}{|c||c|c|c|c|c|} \hline  

i & 2 & 3 & 4 & 5 & 6  \\

\hline \hline
$\gamma_{i}$ (h$^{-1}$) &3.8$\times10^{-2}$ &7.6 $\times 10^{-2}$ & 1.69 $\times 10^{-1}$ &5.063$\times 10^{-2}$& $1.69 \times 10^{-2}$ \\
\hline
$B_{i}$ (h$^{-1}$)& 3.8 $\times10^{-2}$ & 7.6 $\times 10^{-2}$ & 1.69 $\times 10^{-1}$ &5.063$\times 10^{-2}$& $1.69 \times 10^{-2}$ \\
\hline 
$T_{oi} $ &$20$ &20 &20  & 20 & 20\\
\hline

\end{tabular}}{}


\vspace{0.3cm}

{\begin{tabular}{|c|c|c|c|c|c|c|} \hline 

$\alpha_{T_5,T_5}$ & $\alpha_{D,D}$&$\alpha_{T_5,D}$&$\alpha^{\prime}_{T_6,T_6}$& $\alpha^{\prime}_{D,D}$& 
$ \alpha^{\prime}_{T_5,D}$& $A_D$ (h$^{-1}$)\\

\hline \hline
0.1 & 0.8 & 0.1 & 0.1 & 0.8 & 0.1 & 0.1\\
\hline

\end{tabular}}{}

\end{center}

\label{Tab:02} 
\end{table}

As described in Sec. \ref{sec:model}, the only processes that involve changes in the TAC$_{i}$ populations ($i=2,3,...,6$) are their production from a mitosis of a TAC of 
previous generation and their disappearance, since its own mitosis lead to TACs of later generation or DCs. We will consider that the ratios of these 
processes are regulated as follows:

\begin{equation}
L_{T_{i}}(T_{i})=\left( \gamma_{i}+ \frac{B_{i} T_{i}}{(T_{oi}+T_{i})} \right), i=2,...6 \, . \label{eq:22}
\end{equation}

Unfortunately, we do not have experimental determinations on the equilibrium populations for each of the generations of TACs 
and neither about the fractions of cell divisions that originate differentiated cells in generations 4 and 5. This means that 
the parameters we use are not well determined and we have chosen them to satisfy the few known data.
The inclusion of these results is intended to show that for a reasonable choice of parameters, there is a stable fixed point 
for the equations that accounts for the average cell populations and that, moreover, these populations are stable against 
internal noise.

Table \ref{Tab:02} shows the parameters used for calculations under following criteria:
\begin{enumerate}
\item The average total cell population, in homeostatic conditions comes out to 300 cells. This is:

\begin{equation}
S_s+\sum_{i=1}^{i=6} T_{si}+D_{s}=N_{sT} \approx 300 \, .
\end{equation}

\item The average population of DCs ($D_s$) is about 150.

\item The average cell cycle of all TACs is about 10h, that is:

\begin{equation}
 \frac{1}{\sum_{i=1}^{i=6} T_{si}} \left((A_{T_{1}}(T_{s1})+P(S_s)) T_{s1}+\sum_{i=2}^{i=6} L_{T_{i}}(T_{si}) T_{si} \right) \approx \frac{1}{10 \mbox{h}} \,.
\end{equation}
\end{enumerate}

\begin{figure*}[!t]
\centerline{\includegraphics[scale=0.5,clip=true,angle=0]{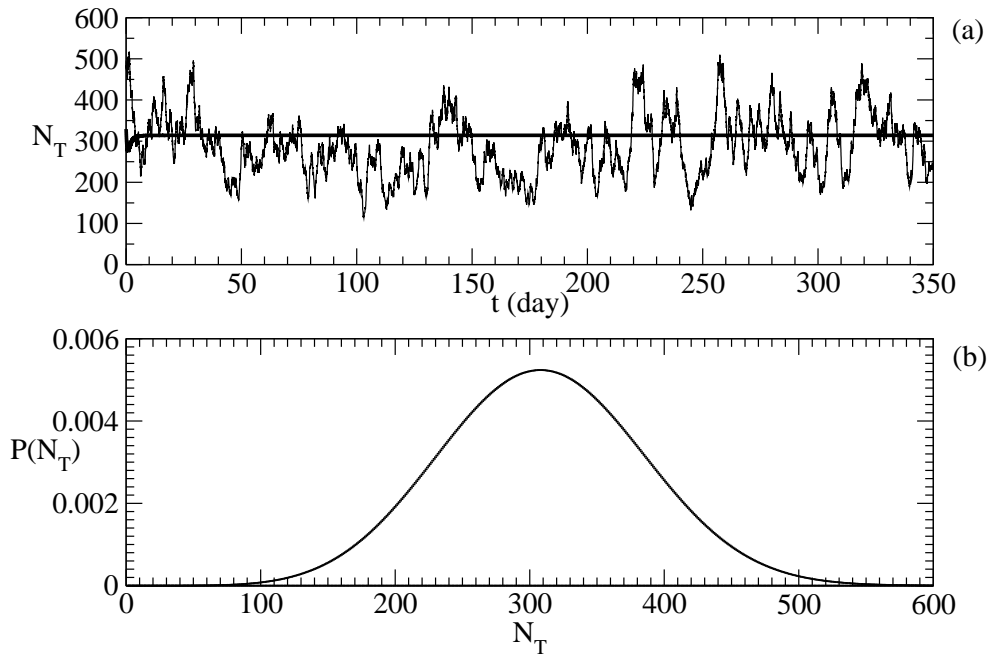}}
\caption{(a) Total cells in the crypt ($N_{T}$) vs. time ($t$). The horizontal line correspond to the solution of the differential equations.
The fluctuating line correspond to a realization of the solution of the master equations.
(b) Relative frequency of values of $N_{T}$ over $10^6$ realizations.}
\label{fig:09}
\end{figure*}

In such conditions the mean cell population in homeostatic conditions results $S_s=15, T_{s1}= 15.62$ (as was already stablished) and 
$T_{s2}=24.54,T_{s3}=24.54,T_{s4}=24.54,T_{s5}=26.88,T_{s6}=24.54, D_{s}=158.41$ (stable fixed point of Eqs.(\ref{eq:02})-(\ref{eq:07})). Figure \ref{fig:08} shows these 
values as well as the population for a realization of master equations solution. On the other hand, Fig.\ref{fig:09}(a) shows the mean population of total cells 
($N_{sT}=314.06$) and, the total cell population for a realization given by the master equations. Figure \ref{fig:09}(b) shows the distribution of the total cells around the 
mean population.

\section{\label{sec:end} Discussion and Conclusions}

In the present work we have developed a compartment model for the evolution of the different cell populations in colonic crypts. The model consists of eight
differential equations, associated with the population of SCs , TACs (six generations) and completely DCs. The equations for SC and TAC1 are coupled and their 
solutions feed the other equations. The need to consider a compartment for each generation of TAC it is related to the fact that these cell states are intermediate and 
transient. Therefore its division necessarily leads to daughter cells qualitative different from their progenitors
(more commited in the process of cell differentiation than their progenitors).

The huge dispersion in the experimental data on the cell populations of the crypts
as well as, their monoclonal character in homeostatic conditions, suggests the need to consider the equations not as differential equations but
as master equations that can be solved using Montecarlo simulations.
Experimental data on evolution times towards monoclonality, suggest
that the cell renewal regimen in homeostasis, is the result of a balance between symmetric divisions of SCs of stochastic character 
that introduces large fluctuations of the cell populations around their mean values.
The model presented here allows to study the possible effects of mutations that can occur
in some of the cells that feed the cell renewal process.
The neutral drift leading to monoclonality can fix the mutant clone or extinguish it. The fixation of the mutant clones lead to atypical crypts, with different 
mean populations, as indication of dysfunctionalities or even the total extinction of the crypt.
As an example of the latter is when a mutant clone without plasticity is fixed. 
The systematic study on the possible effects of different mutations will be the object of future work.

\begin{appendix}

\normalsize

\section{\label{sec:gille} Gillespie Algoritm}

In order to make this article self-contained and save the reader to use Ref. \citen{Gilli}, here it is a brief summary of the 
algorithm as well as some details for this specific case.

Let us consider that the evolution of populations $S$ and $T_1$ are the result of a stochastic process consisting of the occurrence 
of six independent events with a given probability. The events, the probabilities of occurrence and the changes 
that occur in the populations involved are detailed below, when all SCs and TAC$_1$s are equally adapted.

\begin{enumerate}
\item SC-SC Symmetric division of a SC.
\begin{eqnarray}
P_{S,S} \propto M_{S,S} S = a(1) \, , \label{eq:b1}\\
S \rightarrow S+1 \, ,  \nonumber \\
T_1 \rightarrow T_1 \, . \nonumber 
\end{eqnarray}

\item TAC$_1$-TAC$_1$ symmetric division of a SC.
\begin{eqnarray}
P_{S,S} \propto M_{T_{1},T_{1}} S =a(2) \, , \label{eq:b2} \\
S \rightarrow S-1 \, , \nonumber \\
T_1 \rightarrow T_1+2 \, . \nonumber 
\end{eqnarray}

\item SC apoptosis
\begin{eqnarray}
P_{A_S} \propto A_{S} S =a(3) \, , \label{eq:b3} \\
S \rightarrow S-1 \, , \nonumber \\
T_1 \rightarrow T_1 \, . \nonumber 
\end{eqnarray}

\item Plasticity
\begin{eqnarray}
P_{P} \propto P T_{1}=a(4) \, , \label{eq:b4} \\
S \rightarrow S+1, \, , \nonumber \\
T_1 \rightarrow T_1 \, . \nonumber 
\end{eqnarray}

\item TAC$_2$-TAC$_2$ symmetric division of a TAC$_{1}$.
\begin{eqnarray}
P_{TAC_2,TAC_2} \propto N_{T_2} T_{1}=a(5) \, , \label{eq:b5} \\
S \rightarrow S \, , \nonumber \\
T_1 \rightarrow T_{1}-1 \, . \nonumber 
\end{eqnarray}

\item TAC$_1$ apoptosis
\begin{eqnarray}
P_{A_{T_1}} \propto A_{T_{1}} T_{1} =a(6) \, , \label{eq:b6} \\
S \rightarrow S  \, , \nonumber \\
T_1 \rightarrow T_{1}-1 \, . \nonumber 
\end{eqnarray}

\end{enumerate}

Once the event in the step is defined, the clone participating in it is determined by another stochastic process. 
The probability that a cell of the clone $j$ is involved is proportional to $ s_j $ for the SCs and a $ t_j $ for the TAC$_1$.
We discard the probability of simultaneous events and the time between steps $\Delta t$ is calculated from the probability 
density $P(\Delta t)= exp{(-\Delta t \sum_i a(i))}$.

For possible mutations of clone 15, it is assumed that the probability of participation in each event of 
that clone is proportional to $m_{i} s_{15}, (i=1,2,3)$ for SCs and  $m^{\prime}_{i} t_{15}, (i=1,2,3)$, while if $j\neq15$
it remains proportional to $ s_j $ or $ t_j $ . 

If we consider that the regulatory factors of each event (Eqs. \ref{eq:08}-  \ref{eq:14}) do not vary (this depends on the total number of SCs or 
TAC$_1$s), this implies that to include the mutations we must only replace $S$ and $T_1$ by 
$\sum_{j}^{14} s_{j}+m_{i} s_{15}$ in $a(i), (i=1,2,3)$ in Eqs. (\ref{eq:b1}),(\ref{eq:b2}),(\ref{eq:b3}) and  $\sum_{j}^{14} t_{j}+m^{\prime}_{i} t_{15}$ in $a(i)
, (i=4,5,6)$ in Eqs.(\ref{eq:b4}),(\ref{eq:b5}),(\ref{eq:b6}) 
respectively.



\section{\label{Null} Nullclines}

\subsection{\label{subsec:normal} Normal crypts}

Here we show the nullclines corresponding to the equations describing the evolution of SCs and TAC$_1$, Eqs. \ref{eq:00} and \ref{eq:01}
for the case when all cells are identical with the parameters shown in table \ref{Tab:01}. The nullcline corresponding to $dT_{1}/dt=0$  is a simple curve whereas 
that corresponding to  $dS/dt=0$ is multivalued. One of the solutions is a curve that passes through the origin while the other one is similar to a hyperbola, 
with two branches. The region of biological interest is reduced to the first quadrant $(S \geq 0, T_1 \geq 0)$. There are two possible points of equilibrium. One of them is stable 
$(S_s=15, T_{s1}=15.62)$, while the other $(S_s=0, T_{s1}=0)$ is unstable (see Fig.\ref{fig:10}(a)). 

\subsection{\label{subsec:mutant} Crypts with a fixed mutant clone}

As we have seen, in Sec. \ref{sec:Res} , to study the evolution towards monoclonality it is necessary to abandon the continuous description of the differential 
equations. However, once a clone is fixed (whether mutant or not), the differential equations acquire some meaning since they describe the average behavior of 
the populations. If the prevailing clone is normal, the differential equations have the same parameters given by table \ref{Tab:01}.
If instead the mutant clone is fixed, the parameters change according to the effect of the mutation. In particular, if the mutation is such 
that plasticity changes, $P \rightarrow m^{\prime}_{1}P$.

\begin{figure*}[!t]
\centerline{\includegraphics[scale=0.65,clip=true,angle=0]{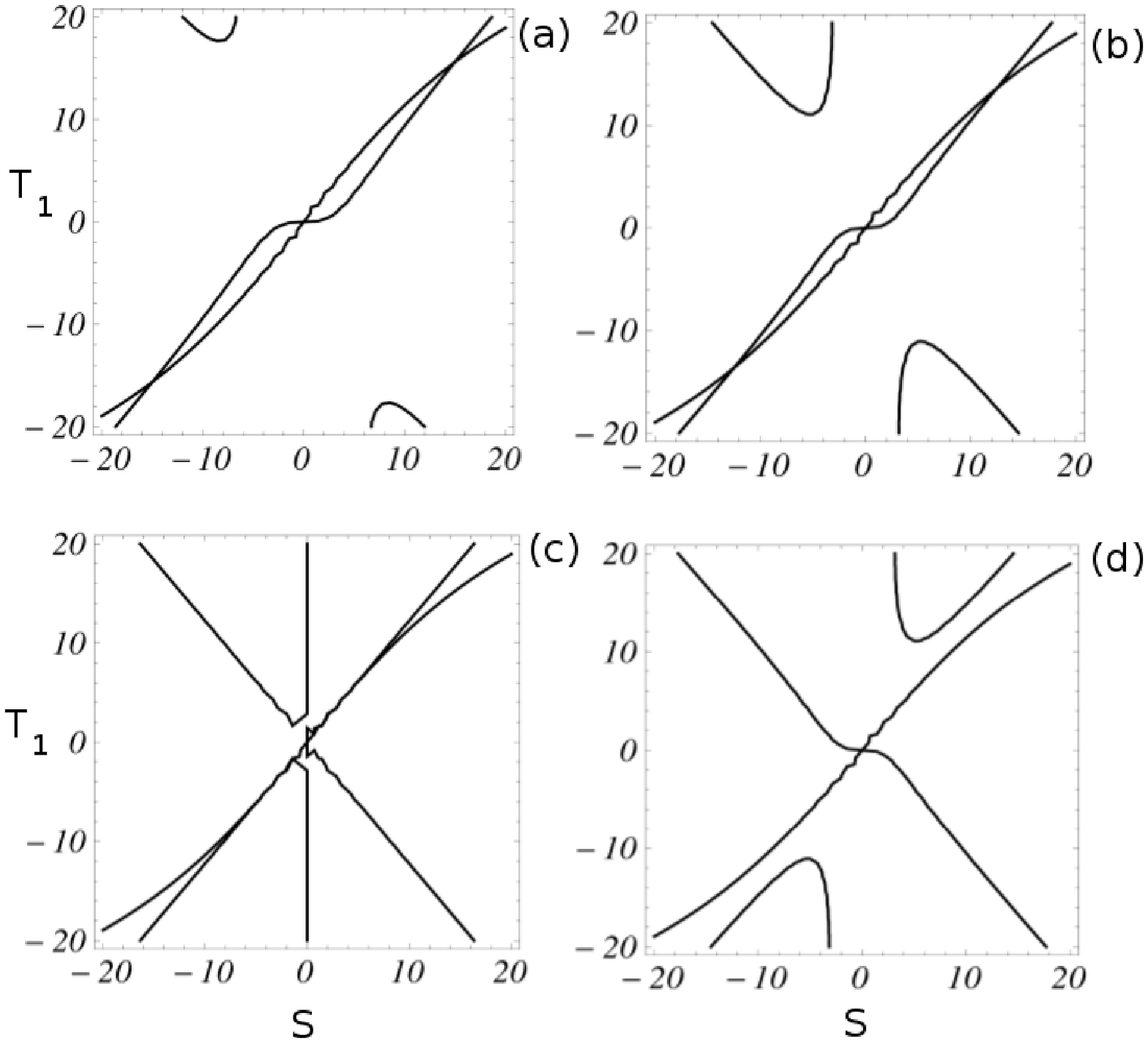}}
\caption{(a) Nullclines for normal crypt. (b) Nullclines for fixed mutant clone with $0<m^{\prime}_{1}<1$.
 (c) Nullclines for $m^{\prime}_{1}=0$.
 (d) Nullclines for $m^{\prime}_{1}<0$}
\label{fig:10}
\end{figure*}

Changes in the parameters modify the nullclines and thus the possible equilibrium points for the populations. The Fig.\ref{fig:10}(b) shows the nullclines corresponding to a 
mutation with decreasing plasticity. For this mutation, the nullcline corresponding to $dT_{1}/dt=0$ does not change but it does to $dS/dt=0$. 
So the values for the equilibrium of populations $(S_s$ and $T_{s1})$ decrease. We can also see how the two branches of the hyperbolic nullcline approach the origin. 
In Fig.\ref{fig:10}(c), we show the limiting case in which the fixed clone has no plasticity. In this case we see how the stable equilibrium point disappears and the 
hyperbolic solution degenerates in its asymptotes. This means if the mutant clone is set, the crypt is extinguished. Mathematically, this case $(P = 0)$ corresponds to a 
bifurcation. For completeness in Fig.\ref{fig:10}(d) we show the nullclines for $P < 0$ (which have no biological sense).

\section{\label{fue} Extinction by fluctuations}

To describe the evolution of populations by means of differential equations, it is not only important to verify the stability of the solutions under the 
parameter fluctuations but also against the fluctuations coming from a stochastic problem of losses and profits. 
If steady-state solutions involve large
populations this may be irrelevant but if it is small colonies (as in the case of crypts), it is mandatory. As an example, we compare the solution of the 
differential equation for the evolution for the SCs population ($N_0$) proposed for a compartment model of crypts \cite{Johnston} and the solution of 
the same through the algorithm shown in the previous appendix.

\begin{figure*}[!t]
\centerline{\includegraphics[scale=0.5,clip=true,angle=0]{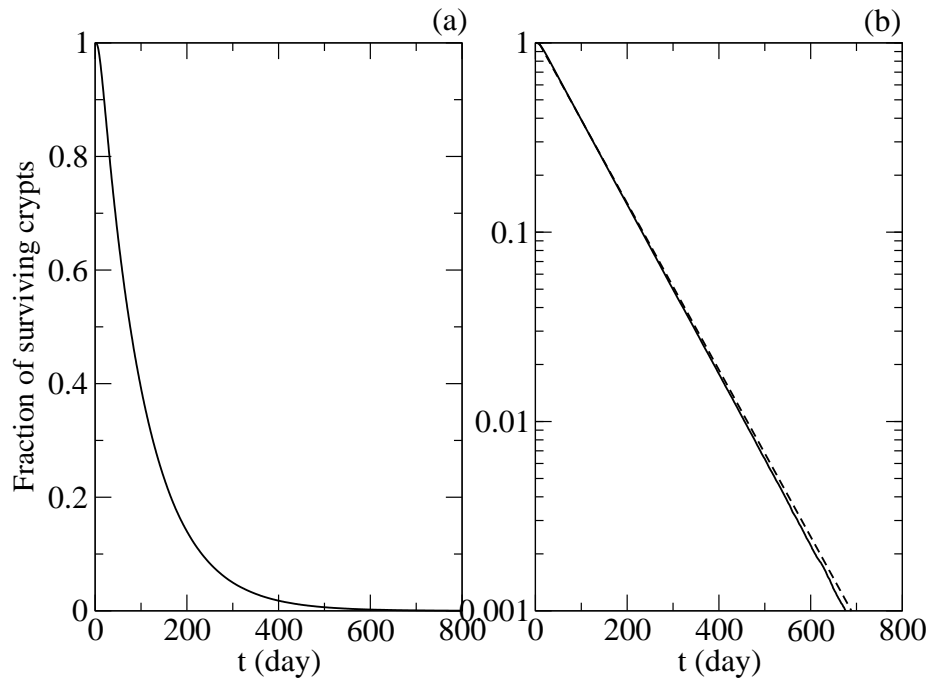}}
\caption{(a) Fraction of surviving crypts vs. time.
(b) The same as in (a) but in log-lin scale. The dashed line correspond to the best exponential fit.}
\label{fig:11}
\end{figure*}

\begin{equation}
\frac{dN_0}{dt}= (\alpha_3-\alpha_1-\alpha_2) N_0- \frac{k_0 N^{2}_{0}}{1+m_{0}N_{0}} \, \, .
\end{equation}
For normal crypts, the authors set $\alpha_1=0.1, \alpha_2=0.3, \alpha_3=0.686, k_0=0.1, m_0=0.1 $. With these parameters the equation has a fixed point at $N_{s0}=4$. 
However when solving the equation with the stochastic algorithm it is clear that the crypts extinguish exponentially with a half life of 98.63 days as can be 
seen in Fig.(\ref{fig:11}). Let us remark that the extinction of crypts is an event that normally occurs rarely \cite{Loeffler}: the extinction rate is much less than $10^{-3}$ day$^{-1}$

\end{appendix}

\section*{Acknowledgements}
This work was partially supported by PIO 144-2014-0100016-CO (CONICET - Argentina).

\end{document}